\documentclass[prl, twocolumn, superscriptaddress, nofootinbib, longbibliography]{revtex4-2}
\usepackage{bm, amsmath, amsfonts, amssymb, mathtools, mathrsfs, braket}
\usepackage{multirow}
\usepackage{graphicx}
\usepackage{float}

\usepackage{comment}
\usepackage[svgnames]{xcolor}
\usepackage[colorlinks,unicode,citecolor=Blue,linkcolor=Blue,urlcolor=Blue,linktocpage,breaklinks=true]{hyperref}

\DeclareMathOperator\Tr{\mathop{\mathrm{Tr}}}

\DeclareMathOperator\spec{\mathop{\mathrm{spec}}}

\usepackage[capitalize,nameinlink]{cleveref} 
\crefname{appendix}{Appendix}{Appendices}
\crefname{section}{Sec.}{Secs.}
\Crefname{section}{Section}{Sections}
\crefname{figure}{Fig.}{Figs.}
\Crefname{figure}{Figure}{Figures}
\crefname{table}{Table}{Tables}
\crefname{equation}{Eq.}{Eqs.}
\Crefname{equation}{Equation}{Equations}
\crefname{definition}{Definition}{Definitions}
\crefname{theorem}{Theorem}{Theorems}
\crefname{proposition}{Proposition}{Propositions}
\Crefname{proposition}{Proposition}{Propositions}
\crefname{proposition*}{Proposition}{Propositions}
\Crefname{proposition*}{Proposition}{Propositions}

\def\titlename{Renormalization of Quantum Operations: Parity-Time Transition and Chaotic Flows}

\begin{document}

\title{\titlename}

\author{Atsushi Oyaizu}
\email{oyaizu@cat.phys.s.u-tokyo.ac.jp}
\affiliation{Department of Physics, University of Tokyo, 7-3-1 Hongo, Bunkyo-ku, Tokyo 113-0033, Japan}

\author{Hongchao Li}
\affiliation{Department of Physics, University of Tokyo, 7-3-1 Hongo, Bunkyo-ku, Tokyo 113-0033, Japan}

\author{Masaya Nakagawa}
\affiliation{Department of Physics, University of Tokyo, 7-3-1 Hongo, Bunkyo-ku, Tokyo 113-0033, Japan}

\author{Masahito Ueda}
\affiliation{Department of Physics, University of Tokyo, 7-3-1 Hongo, Bunkyo-ku, Tokyo 113-0033, Japan}
\affiliation{Fundamental Quantum Science Program (FQSP), TRIP Headquarters, RIKEN, Wako 351-0198, Japan}
\affiliation{Institute for Physics of Intelligence, University of Tokyo, 7-3-1 Hongo, Bunkyo-ku, Tokyo 113-0033, Japan}

\date{\today}

\begin{abstract}
  The renormalization group (RG) in statistical physics focuses on ground-state properties of equilibrium systems. However, it is unclear how it should be generalized to nonunitary quantum dynamics caused by dissipation and measurement backaction, in which the notion of conserved energy is absent. Here, we extend the RG to cover nonunitary quantum dynamics governed by quantum operations. By performing coarse-graining in real time, we find that the competition between decoherence and coherent dynamics plays a decisive role in the behavior of the RG flow. In particular, we find that chaotic behavior without fixed points emerges in the RG flow when coherent dynamics is dominant, with the parity-time transition serving as a prototypical example. The measurement-induced parity-time transition belongs to the universality class of the one-dimensional Yang-Lee edge singularity, which serves as a guide for experimentally realizing imaginary fields in lattice spin systems with a quantum system.
\end{abstract}

\maketitle

\textit{Introduction}.---\!
Recent years have witnessed significant interest in the dynamics of open quantum systems subject to dissipation and measurement backaction~\cite{BreuerPetruccione_2007, daley2014quantum, WisemanMilburn}, and various nonequilibrium critical phenomena and universality have been discovered.
Examples include real-complex spectral transitions in non-Hermitian systems~\cite{lee1952statistical, Bergholtz_2021, Ashida_2020, bender2024pt}, measurement-induced entanglement transitions~\cite{skinner2019measurement, li2018quantum, chan2019unitary, gullans2020dynamical, fisher2023random}, and dissipative phase transitions intrinsic to open quantum systems~\cite{kessler2012dissipative, Minganti_2018, haga2023quasiparticles}.
These diverse critical phenomena can be understood as arising from the competition among decoherence, dissipation, and coherent quantum dynamics.

The renormalization group (RG) has successfully unveiled phases of matter and phase transitions at thermal equilibrium~\cite{kadanoff1966scaling, wilson1971renormalization, wilson1971renormalization2, wilson1974renormalization, wilson1975renormalization}.
Extensions to nonunitary systems have been actively developed, for instance, within the framework of Keldysh field theory~\cite{Sieberer_2016, kamenev2023field}, where the RG is formulated for continuum quantum fields.
An outstanding question is how to construct an RG directly at the level of quantum operations in discrete time.
While recent progress has been made in formulating real space RG on mixed states~\cite{sang2024mixed}, its roles in nonunitary quantum dynamics remain largely unexplored.
 
In this Letter, we construct and examine an RG scheme that directly acts on quantum operations and performs an RG transformation in real time.
We first demonstrate that the parity-time (PT) transition~\cite{Bender_1998}, a paradigmatic dynamical phase transition in open quantum systems~\cite{bender2024pt, nakanishi2025lindbladian}, of a single qubit can be captured by a \textit{chaotic} transition of the associated RG flow, at which the associated fixed points disappear.
We then extend this result to generic finite-level systems, revealing a competition between measurements and unitary dynamics that governs the presence or absence of fixed points.
We uncover that the PT transition belongs to the universality class of the Yang-Lee edge singularity~\cite{lee1952statistical, bena2005statistical, fisher1978yang, kurtze1979yang, fisher1980yang, cardy1985conformal, cardy2024yang, li2025yang, li2023yang}, which can be used to experimentally implement imaginary magnetic fields in lattice spin systems.
We also show that, by explicitly taking the purified ancilla qudits into account, the RG transformation of nonunitary dynamics can be considered as that of the ancillas in a matrix product state~\cite{verstraete2005renormalization, cirac2017matrix}.

\textit{Renormalization of nonunitary quantum dynamics}.---\!
We consider nonunitary discrete-time evolution of a quantum system represented by the quantum operation $\Phi(\cdot)\coloneqq \sum_{m}K_{m}(\cdot)K_{m}^{\dagger}$ as
\begin{equation}\label{eq:quantum-operation}
  \rho \to \Phi(\rho) \to \Phi^2(\rho) \to \cdots,
\end{equation}
where a set of Kraus operators $\left\{ K_{m} \right\}_{m}$ satisfy the trace non-increasing condition $\sum_{m}K_{m}^{\dagger}K_{m}\leq \mathbf{1}$. 
When the trace-preserving condition $\sum_{m}K_{m}^{\dagger}K_{m}=\mathbf{1}$ is imposed, $\Phi$ becomes a quantum channel that describes nonunitary processes such as dissipation; otherwise, $\Phi$ represents the evolution of a state  subject to a sequence of measurements with a postselection of measurement outcomes~\cite{WisemanMilburn, Nielsen_Chuang_2010}.
As a special case of \cref{eq:quantum-operation}, we also consider the evolution of a pure state under a postselected quantum trajectory,
\begin{equation}\label{eq:non-Hermitian}
  \ket{\psi} \to K\ket{\psi} \to K^2\ket{\psi} \to \cdots,
\end{equation}
where $K$ is a Kraus operator corresponding to a particular measurement outcome.
The normalization of states is not explicitly written in \cref{eq:quantum-operation,eq:non-Hermitian}.

Under this setting, we investigate the long-time behavior of the dynamics by applying Kadanoff's block-spin transformation~\cite{kadanoff1966scaling} on a quantum operation.
Let $\Phi[\bm{g}]$ be a generic quantum operation parametrized by $\bm{g}$.
By blocking $b$ discrete-time steps, we find a set of renormalized coupling constants $\bm{g}'$ from
\begin{equation}\label{eq:Phi-RG}
  \Phi[\bm{g}'] \propto (\Phi[\bm{g}])^{b}.
\end{equation}
The same transformation can also be applied to a Kraus operator $K$ in \cref{eq:non-Hermitian} as
\begin{equation}\label{eq:K-RG}
  K[\bm{g}'] \propto (K[\bm{g}])^{b}.
\end{equation}
Repeated applications of the RG transformation generate the RG flow $\bm{g}_{0} \to \bm{g}_{1} \to \bm{g}_{2} \to \cdots$, where $\bm{g}_{n+1}$ is determined recursively from $\bm{g}_{n}$ via \cref{eq:Phi-RG} or \cref{eq:K-RG}.
Our RG scheme concerns coarse-graining in real time and is distinct from existing formulations of RG for non-Hermitian Hamiltonians in Refs.~\cite{ashida2017parity, nakagawa2018non, li2023yang, burke2025non, yang2026asymptotic, yamamoto2026complex, tajima2026non}.

\textit{Chaotic transition of the RG flow}.---\!
As a minimal example, we consider a single-qubit system evolving under a unitary $U=\exp(\mathrm{i}h\sigma_{z})$ with $h\in [0, \pi/2]$ and repeated weak measurements of the $x$ component of its spin with a measurement operator
\begin{equation}\label{eq:Mm}
  M_{m} = \frac{1}{\sqrt{2\cosh 2\Gamma}} \exp\left(m\Gamma \sigma_{x}\right), \quad m\in \left\{ \pm 1 \right\},
\end{equation}
where $\sigma_{z}$ and $\sigma_{x}$ are the Pauli matrices, and $\Gamma\geq 0$ represents the strength of the measurement with $\Gamma=\infty$ and $0$ corresponding to a projective measurement and  no measurement, respectively.
We define the Kraus operator $K_{m}\coloneqq \sqrt{U}M_{m}\sqrt{U}$ and postselect a sequence of measurement outcomes $\bm{m}=(+1, \cdots, +1)$.
Then, the state evolution is given by \cref{eq:non-Hermitian} with $K\coloneqq K_{m=+1}$, and \cref{eq:K-RG} with $\bm{g}\coloneqq (\Gamma,h)$ determines the RG equations.
Specifically, by taking $b=2$, they are given as
\begin{align}
  \tanh \Gamma' &= \frac{\cos h \sinh 2\Gamma}{\left| \mathrm{e}^{\mathrm{i}h}\cosh^2 \Gamma + \mathrm{e}^{-\mathrm{i}h}\sinh^2 \Gamma \right|}, \label{eq:RG-Gamma}\\
  \mathrm{e}^{2\mathrm{i}h'} &= \mathrm{e}^{2\mathrm{i}h}\frac{\mathrm{e}^{\mathrm{i}h}\cosh^2 \Gamma + \mathrm{e}^{-\mathrm{i}h}\sinh^2 \Gamma}{\mathrm{e}^{-\mathrm{i}h}\cosh^2 \Gamma + \mathrm{e}^{\mathrm{i}h}\sinh^2 \Gamma}. \label{eq:RG-h}
\end{align}

\begin{figure}[tpb]
  \centering
  \includegraphics[width=\linewidth]{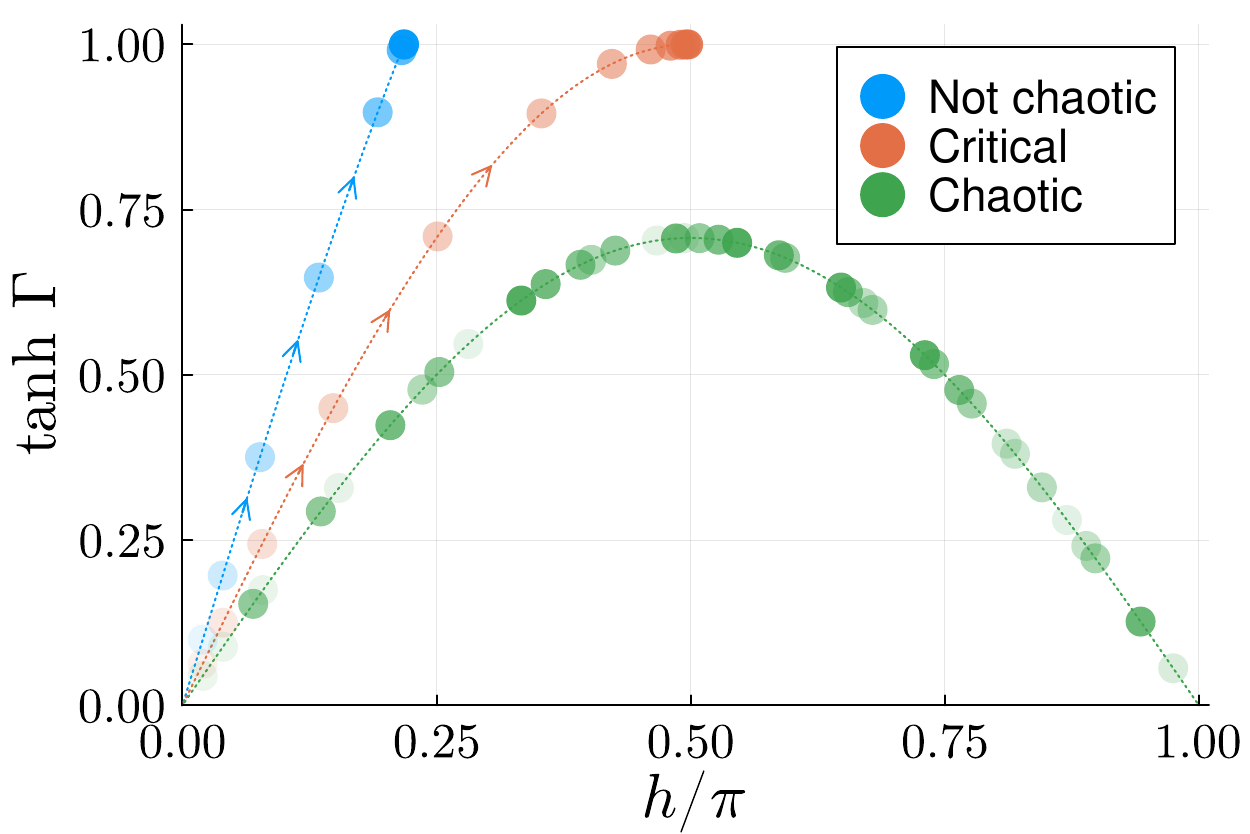}
  \caption{
    RG flows, determined from \cref{eq:RG-Gamma,eq:RG-h}, are shown for three different choices of initial parameters.
      The dots represent the renormalized coupling constants $\bm{g}_{n}=(\Gamma_{n},h_{n})$ after $n$ applications of the RG transformation in \cref{eq:K-RG}, with darker colors corresponding to larger $n$.
      In the non-chaotic phase (blue) with $h<h_{\mathrm{c}}(\Gamma)$, the RG flow converges monotonically to a fixed point at $\tanh \Gamma = 1$, as indicated by arrows.
      In the chaotic phase (green) with $h>h_{\mathrm{c}}(\Gamma)$, the RG flow does not converge to a fixed point nor undergoes a limit cycle.
      The critical line (orange) corresponds to $h = h_{\mathrm{c}}(\Gamma)$.
      The initial parameters $(\Gamma_{0}, h_{0})$ are chosen as $(0.1, 0.02\pi)$, $(0.06, 0.02\pi)$, and $(0.04, 0.02\pi)$ for the non-chaotic, critical, and chaotic phases, respectively.
  }
  \label{fig:phase}
\end{figure}

Notably, these RG equations have no (stable) fixed point once $h$ exceeds the critical value $h_{c}$.
In fact, \cref{eq:RG-Gamma,eq:RG-h} can be recast~\cite{dolan1995chaotic} in terms of the following logistic map for a single parameter,
\begin{equation}\label{eq:logistic}
  x' = 4x(1-x),
\end{equation}
where $x\coloneqq 1-\cosh^2 \Gamma \cos^2 h$.
When $|h|<h_{\mathrm{c}}(\Gamma)\coloneqq \arcsin \tanh \Gamma$, $x<0$ and hence the RG equation flows to a fixed point at $x = -\infty$.
However, when $|h|>h_{\mathrm{c}}(\Gamma)$, $0<x<1$, and the RG equation in \cref{eq:logistic} becomes a chaotic map with no fixed points.\footnote{Regardless of the value of the number $b$ of coarse-grained blocks, the resulting RG equation becomes a chaotic map; see Supplemental Material~\cite{supplement} for details.}
Hence, $h = h_{\mathrm{c}}(\Gamma)$ marks the critical point at which the RG flow becomes chaotic (see \cref{fig:phase}.)

Physically, the presence of a chaotic RG flow indicates a breakdown of the UV-IR separation principle~\cite{Derrida_1983, damgaard1991chaotic, Damgaard:1991zb, supplement}, on which the conventional RG is based~\cite{wilson1974renormalization}.
Since this principle applies to most stable equilibrium phases of matter, the convergence of the RG flow has been assumed in many cases.
However, as Wilson pointed out~\cite{Wilson:1971ag, wilson1971renormalization, wilson1974renormalization}, the convergence of the flow is actually a nontrivial problem (see also Refs.~\cite{mckay1982spin,  svrakic1982hierarchical, Derrida_1983, berker1984hierarchical, banavar1987chaos, damgaard1991chaotic, Damgaard:1991zb, dolan1995chaotic, dolan2002one, glazek2002limit, morozov2003can, curtright2012renormalization, ilderton2020renormalization, jiang2021chaotic, bosschaert2022chaotic, calero2023chaotic}).
Our example here demonstrates that one cannot assume the convergence a priori in nonequilibrium dynamics.

\textit{PT transition as a chaotic transition}.---\!
This chaotic transition of the RG flow at $h = h_{\mathrm{c}}(\Gamma)$ can be understood as a consequence of spontaneous breaking of PT symmetry~\cite{Bender_1998, Bender_2007, bender2024pt} of the Kraus operator, defined as $\sigma_{x}K^{*}\sigma_{x} = K$.
Because of this symmetry, the two eigenvalues $\lambda_{1, 2}$ of $K$ are either both real or form a complex conjugate pair.
A direct computation shows that $\lambda_{1,2}\in \mathbb{R}$ and the PT symmetry is unbroken in the non-chaotic phase, whereas
$\lambda_{1}=\lambda_{2}^{*}$ and the PT symmetry is spontaneously broken in the chaotic phase.
At the critical point $h = h_{\mathrm{c}}(\Gamma)$, the two eigenvalues coalesce and $K$ becomes non-diagonalizable.
As we shall see below, this PT transition is caused by a competition between the measurement backaction and the unitary dynamics, whose strengths are quantified by $\Gamma$ and $h$, respectively.

The condition $|h|<h_{\mathrm{c}}(\Gamma)$ for the non-chaotic phase is satisfied when the measurement strength $\Gamma$ is sufficiently large.
Here, the fixed point at $x=-\infty$ corresponds to $\Gamma=\infty$, which represents a projective measurement, as
can easily be understood from the relation $(\sqrt{U}P_{+}\sqrt{U})^2 \propto \sqrt{U}P_{+}\sqrt{U}$, where $P_{+} =\ket{+}\!\bra{+}\coloneqq\lim_{\Gamma \to \infty}M_{m}$ is a projection operator.
As such a fixed point is approached, the measurement strength $\Gamma$ increases monotonically.

In the chaotic phase with $|h|>h_{\mathrm{c}}(\Gamma)$, on the other hand, the unitary dynamics dominate the measurement backaction.
We now have a clear physical picture behind the chaotic transition:
Measurements tend to create an RG fixed point, while a unitary evolution prevents reaching it.
It seems, however, rather peculiar that a unitary evolution alone causes a chaotic behavior in the RG flow.
To understand this point, let us recall that chaotic maps are those that (i) are deterministic (i.e., no randomness is involved in the map), (ii) have exponential sensitivity to initial parameters, and (iii) have no (quasi-)periodic orbits~\cite{strogatz2024nonlinear}.
In our case, the condition (i) is satisfied due to the postselection of measurement outcomes.
The condition (ii) is attributed to the persistent oscillatory behavior of the state in the PT-broken phase.
Since the state never converges to a steady state, an infinitesimal perturbation in the initial system parameters affects the period of oscillations, resulting in an $O(1)$ difference in the states at late times~\cite{supplement}.
The condition (iii) is due to the fact that the oscillation period is incommensurate with the discrete-time unit (which we set to be unity) that defines the renormalization scale.
This fact hinders the flow to reach the fixed point at $x = 0, 1$~\cite{supplement}.

\textit{Criticality}.---\!
Let us now examine the critical behavior as $h$ approaches $h_{\mathrm{c}}(\Gamma)$ from below.
To this end, we consider a quantity
\begin{equation}\label{eq:mu-def}
  \mu(h,L_{1},L_{2}) \coloneqq \frac{\Tr(K^{L_{1}}\sigma_{z}K^{L_{2}}\rho K^{\dagger}{}^{L_{2}}\sigma_{z}K^{\dagger}{}^{L_{1}})}{\Tr\left( K^{L_{1}+L_{2}} \rho K^{\dagger}{}^{L_{1}+L_{2}} \right)},
\end{equation}
which can be observed in the measurement dynamics~\cite{supplement}.
Here, $\rho$ is an arbitrary initial state.
A straightforward calculation shows that as $h$ approaches $h_{\mathrm{c}}(\Gamma)$ from below, 
\begin{equation}
  \lim_{L_{1,2} \to \infty}\mu(h,L_{1},L_{2}) \propto (h_{\mathrm{c}}-h)^{2\sigma},
\end{equation}
where $\sigma = -1/2$ is the critical exponent that characterizes the Yang-Lee edge singularity in one dimension~\cite{fisher1980yang};
see Supplemental Material~\cite{supplement} for details.
Here, the Yang-Lee universality emerges from the fact that this measurement model can be mapped to the imaginary-time evolution of the classical one-dimensional Ising chain subject to a pure imaginary magnetic field~\cite{supplement}.
The mapping is achieved through identification of the Kraus operator $K$ with the transfer matrix of the Ising model.
The chaotic transition of the RG flow discussed above is then mapped to the Yang-Lee edge singularity~\cite{lee1952statistical, bena2005statistical, fisher1978yang, kurtze1979yang, fisher1980yang, cardy1985conformal, cardy2024yang, li2025yang, li2023yang}, at which the transfer matrix has a PT-transition~\cite{matsumoto2022embedding, meisinger2013pt, supplement}.
Thus, our result also gives a clear physical understanding of the chaotic RG flows in the Yang-Lee edge singularity~\cite{dolan1995chaotic, dolan2002one}.

\textit{Conditions for the emergence of chaotic RG flows.}---\!
The above argument for the emergence of a chaotic flow can be generalized to a generic finite-level system (i.e., a qudit), yielding the following proposition:

\textbf{Proposition.}
Consider a finite-level system and its nonunitary dynamics in \cref{eq:quantum-operation,eq:non-Hermitian}, and the associated RG equations in \cref{eq:Phi-RG,eq:K-RG}.
Let $\lambda_{1}, \lambda_{2},\cdots$ be the eigenvalues of $K[\bm{g}]$ (or $\Phi[\bm{g}]$) with decreasing order in magnitudes, i.e., $|\lambda_{1}|\geq |\lambda_{2}|\geq \cdots$, and $\Delta \coloneqq |\lambda_{1}|-|\lambda_{2}|$ be the damping gap.
Then, a chaotic RG flow of the above type emerges if (a) $\Delta = 0$ and (b) $\arg(\lambda_{1}/\lambda_{2})\notin \pi \mathbb{Q}$ are both satisfied, where $\mathbb{Q}$ denotes a set of rational numbers.

The proof goes as follows.
The long-time behavior of the dynamics is dominated by a few eigenstates corresponding to eigenvalues $\lambda_{1},\lambda_{2},\cdots$ with the largest modulus.
Let $\theta\coloneqq \arg(\lambda_{1}/\lambda_{2})$ and $x\coloneqq \sin^2 \theta$.
Then, under the RG transformation in \cref{eq:K-RG,eq:Phi-RG} with $b=2$, $\theta \to 2 \theta\mod 2\pi$, i.e., $x \to \sin^2 (2\theta) = 4x(1-x)$, which is nothing but \cref{eq:logistic}.
Since $\theta\notin \pi \mathbb{Q}$, this map is chaotic.
Here, the conditions $\Delta=0$ and $\theta\notin \pi \mathbb{Q}$ account for the sensitivity and the incommensurability, respectively.
This completes the proof.

The two conditions (a), (b) above completely characterize the emergence of a chaotic RG of the above type.
As can be checked directly, in the PT transition considered in the previous section, the two conditions are satisfied in the PT-broken (chaotic) phase, while the gapless condition (a) is violated in the PT-unbroken (non-chaotic) phase as $\Delta>0$ .

\textit{RG of dynamics without postselection}.---\!
As a special case of the above argument, let us consider the trace-preserving dynamics that requires no postselection, in which $\Phi$ in \cref{eq:quantum-operation,eq:Phi-RG} is a quantum channel denoted by $\mathcal{E}$.
Importantly, by utilizing the well-known equivalence between a quantum channel and a matrix product state (MPS)~\cite{perez2006matrix}, \cref{eq:Phi-RG} can also be interpreted as an RG transformation for the MPS formed by $\left\{ K_{m} \right\}$~\cite{verstraete2005renormalization, cirac2017matrix},
which has been instrumental in the classification of gapped ground states in one spatial dimension~\cite{schuch2011classifying, chen2011classification}.

\begin{figure}[tpb]
  \centering
  \includegraphics[width=.5\linewidth]{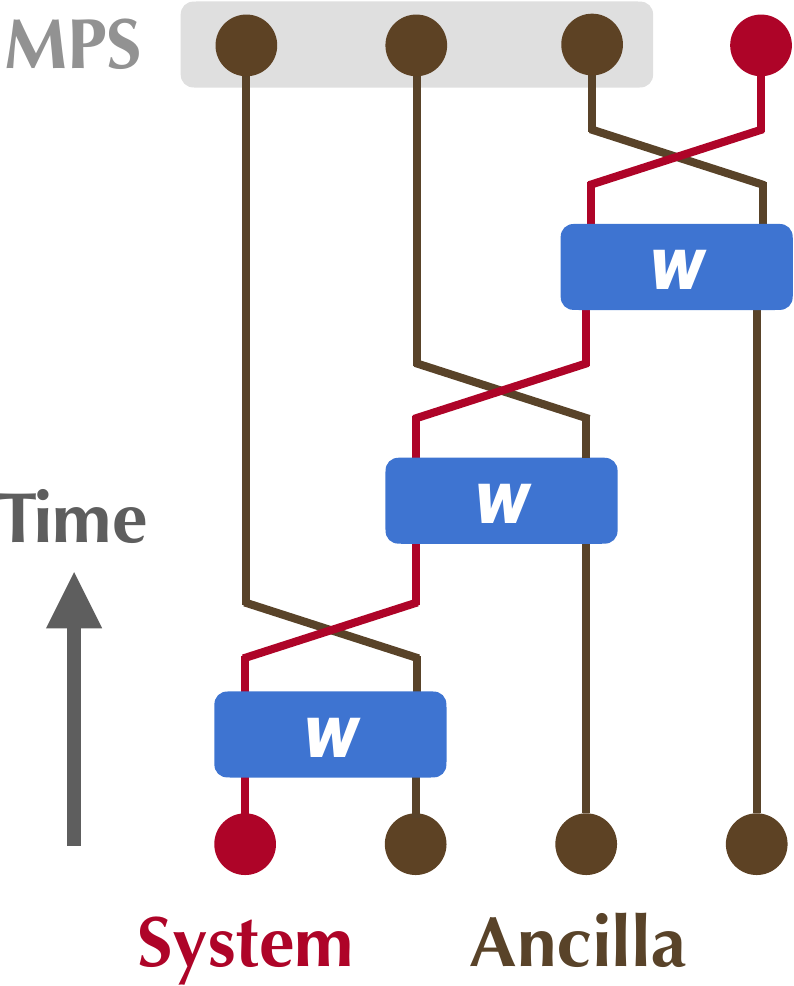}
  \caption{
    A sequential quantum circuit, in which the system qudit (red) sequentially interacts with the rest of the ancilla qudits.
    Each interaction is described by a unitary $W$.
    The system qudit undergoes a nonunitary evolution, while the ancilla qudits after the interactions are described by an MPS.
  }
  \label{fig:SQC}
\end{figure}

To make this correspondence explicit, let us consider the sequential quantum circuit~\cite{schon2005sequential, schon2007sequential, banuls2008sequentially, wei2022sequential} shown in \cref{fig:SQC}, in which a system qudit at the leftmost site interacts sequentially with the rest of the $L$ ancilla qudits.
The interaction between the system qudit and one of the ancilla qudits prepared in a state $\ket{\sigma}$ is described by a two-qudit unitary operator $W$,
\begin{equation}\label{eq:W-single}
  W (\ket{\psi} \otimes \ket{\sigma}) = \sum_{m}(K_{m}\ket{\psi}) \otimes \ket{m},
\end{equation}
where $\ket{\psi}$ is an input state of the system and $K_{m}\coloneqq (\mathbf{1} \otimes \bra{m}) W (\mathbf{1}\otimes \ket{\sigma})$ is the Kraus operator acting on it;
the operators $\left\{ K_{m} \right\}$ satisfy the completeness condition thanks to the unitarity of $W$.
In other words, $W$ gives the purification of the channel $\mathcal{E} = \sum_{m}K_{m}(\cdot)K_{m}^{\dagger}$.
It then follows from \cref{eq:W-single} that the sequential interactions shown in \cref{fig:SQC} maps an initial state $\ket{\psi_{\mathrm{i}}} \otimes \ket{\sigma}^{\otimes L}$ to another state
\begin{equation}\label{eq:sequential-final}
  \sum_{\left\{ m_{i} \right\}} \left(K_{m_{L}}\cdots K_{m_{1}}\ket{\psi_{\mathrm{i}}}\right) \otimes \ket{m_{L},\cdots, m_{1}}.
\end{equation}
By projecting the final state of the system onto some state $\ket{\psi_{\mathrm{f}}}$, we obtain an MPS $\ket{\Psi}=\sum_{\bm{m}}\braket{\psi_{\mathrm{f}}|K_{m_{L}}\cdots K_{m_{1}}|\psi_{\mathrm{i}}}\ket{m_{L},\cdots,m_{1}}$ as the ancilla state after all the interactions.

\Cref{eq:Phi-RG} (with $\Phi$ replaced by $\mathcal{E}$) can be interpreted either as an RG transformation of the nonunitary dynamics of the system, or as that of the ancilla state in an MPS;
the RG scheme for the latter is formulated in Refs.~\cite{verstraete2005renormalization, cirac2017matrix},
in which $\mathcal{E}$ is the transfer matrix defined for an MPS.
Hence, the renormalization of the nonunitary dynamics can be achieved by coarse-graining the ancilla qudits.

In the RG theory for MPSs, the convergence of RG flows is ensured as long as one focuses on an MPS that is not a superposition of macroscopically distinct states (i.e., a cat state)~\cite{verstraete2005renormalization, cirac2017matrix}.
Such a condition translates into the \textit{irreducibility} of the corresponding channel $\mathcal{E}$, which asserts that there does not exist a nontrivial Hermitian projector $P\neq 0,\mathbf{1}$ such that $\mathcal{E}(P\mathcal{M}_{D}P)\subseteq P\mathcal{E}(\mathcal{M}_{D})P$, where $\mathcal{M}_{D}$ denotes a set of $D\times D$ complex matrices on which $\mathcal{E}$ acts~\cite{Wolf2012Lecture}.
For an irreducible $\mathcal{E}$, it is known that the peripheral spectrum $S\coloneqq\spec \mathcal{E}\cap \mathrm{e}^{\mathrm{i}\mathbb{R}}$ satisfies $S=\left\{\exp(2\pi \mathrm{i} k/p)\right\}_{k\in\mathbb{Z}_{p}}$ for $p\in\{1,2,\dots,D^{2}\}$~\cite{fannes1992finitely, perez2006matrix, Wolf2012Lecture}. Consequently, $\arg(\lambda_{1}/\lambda_{2})\in\pi\mathbb{Q}$ for any $\lambda_{1},\lambda_{2}\in S$, and the condition (b) of the above proposition cannot be satisfied.  
Hence, if $\mathcal{E}$ is irreducible, the RG equation~\eqref{eq:Phi-RG} cannot exhibit chaotic flows and must converge to fixed points.

Specifically, if $\mathcal{E}$ has $S=\left\{ 1 \right\}$ (no degeneracy), then every RG flow converges monotonically to a fixed point.
In this case, the fixed-point form of the Kraus operators is given as~\cite{verstraete2005renormalization}
\begin{equation}
  K^{\text{fixed}}_{m = (n,\alpha)} = \sqrt{\lambda_{\alpha}}\ket{\lambda_{\alpha}}\!\bra{n},
\end{equation}
where $\lambda_{\alpha}\in [0,1]$ and $\ket{\lambda_{\alpha}}$ specifies a decomposition of the unique steady state of $\mathcal{E}$ as $\rho_{\mathrm{ss}} = \sum_{\alpha}\lambda_{\alpha}\ket{\lambda_{\alpha}}\!\bra{\lambda_{\alpha}}$, and $\left\{ \ket{n} \right\}$ is an arbitrary set of orthonormal basis.
In quantum measurement theory, these Kraus operators represent a complete (in particular, sharp) measurement~\cite{WisemanMilburn}, where a single measurement completely collapses the input state to $\ket{\lambda_{\alpha}}\!\bra{\lambda_{\alpha}}$.
If a measurement is complete, then all pieces of information about the input state are lost from the system; away from the fixed point, such a loss of information is incomplete.
In Ref.~\cite{Zamolodchikov:1986gt}, the author makes an analogy between the monotonicity of RG flows and the time evolution of dissipative systems.
Our formulation here gives a concrete physical picture to this observation, where the irreversibility of a flow (i.e., convergence to a fixed point) indeed originates from dissipation.

On the other hand, as long as we consider the RG of quantum channels per se, one is not restricted to irreducible channels.
If a channel is \textit{reducible}, the convergence to a fixed point is not guaranteed, which may result in a chaotic flow.
As an example, let us consider a quantum channel that has a decoherence-free subspace (DFS)~\cite{palma1996quantum, zanardi1997noiseless, duan1997preserving, lidar1998decoherence, lidar1999concatenating}.
In this case, all the Kraus operators take a block-diagonal form~\cite{lidar1999concatenating}:
\begin{equation}\label{eq:Km-DFS}
  K_{m} = \begin{pmatrix} g_{m}U & \mathbf{0} \\ \mathbf{0} & K'_{m} \end{pmatrix} = g_{m}U \oplus K_{m}',
\end{equation}
where $g_{m}\in \mathbb{C}$ with $\sum_{m\in \mathcal{M}}|g_{m}|^2 = 1$ and $U$ is a unitary operator.
The corresponding channel $\mathcal{E}=\sum_{m}K_{m}\otimes K_{m}^{*}$ is manifestly reducible, since one can take $P$ to be the projector onto the DFS.
From \cref{eq:Km-DFS}, $\mathcal{E}$ has a block-diagonal structure which contains $U \otimes U^{*}$.
This means that the channel $\mathcal{E}$ involves $\mathrm{e}^{\mathrm{i}(\theta_{i}-\theta_{j})}$ in its spectrum,
where $\mathrm{e}^{\mathrm{i}\theta_{i}}$ and $\mathrm{e}^{\mathrm{i}\theta_{j}}$ are the eigenvalues of $U$.
Since  $U$ is a generic unitary matrix, there generally exist $i$ and $j$ such that $\theta_{i}-\theta_{j}\notin \pi \mathbb{Q}$.
Thus, $\mathcal{E}$ satisfies the conditions of the above proposition and therefore exhibits a chaotic RG flow.

\textit{Discussions}.---\!
In this Letter, we present the RG transformation of discrete-time nonunitary quantum dynamics and show that the competition between measurements/dissipation and unitary dynamics gives rise to an unconventional chaotic transition of the associated flow.
Here, nonunitary processes such as measurements and dissipation drive the flow toward an RG fixed point, whereas unitary dynamics prevents the flow from approaching it.
It is worthwhile to generalize our formulation to higher-dimensional systems with many-body interactions, where dissipative discrete-time crystalline order~\cite{Gong:2018yfa, gambetta2019discrete, riera2020time} and its extension~\cite{haga2026time} may admit a similar characterization in terms of a chaotic RG flow.
Furthermore, it is also a promising direction to extend our RG scheme to quantum trajectory dynamics, where the presence of random measurement backaction in the dynamics makes the analysis challenging.

\begingroup
\renewcommand{\addcontentsline}[3]{}
\begin{acknowledgments}
  \textit{Acknowledgements.}---\!
  A.O. thanks Akihiro Hokkyo and Kenji Shimomura for helpful discussions, and Norifumi Matsumoto for early contributions to this project.
  A.O. and H.L. are
  supported by Forefront Physics and
  Mathematics Program to Drive Transformation (FoPM),
  a World-Leading Innovative Graduate Study (WINGS)
  Program, the University of Tokyo.
  H.L is also supported by JSPS KAKENHI Grant No. JP24KJ0824.
  M.N. is supported by JSPS KAKENHI Grant No. JP24K16989. 
  M.U. is supported by JSPS KAKENHI Grant No. JP22H01152 and the CREST program ``Quantum Frontiers'' of JST (Grant No. JPMJCR23I1)
  and the RIKEN TRIP initiative.
\end{acknowledgments}
\endgroup

\let\oldaddcontentsline\addcontentsline
\renewcommand{\addcontentsline}[3]{}
\bibliography{../ref}
\let\addcontentsline\oldaddcontentsline


\clearpage
\onecolumngrid

\setcounter{secnumdepth}{3}

\renewcommand{\theequation}{S\arabic{equation}}
\renewcommand{\thefigure}{S\arabic{figure}}
\renewcommand{\thetable}{S\arabic{table}}

\setcounter{equation}{0}
\setcounter{figure}{0}
\setcounter{table}{0}
\setcounter{section}{0}
\setcounter{tocdepth}{0}

\numberwithin{equation}{section}

\begin{center}
{\bf \large Supplemental Material for \\
``\titlename''}
\end{center}

\tableofcontents

\section{Calculation of the Lyapunov exponent}
\label{asec:Lyapunov}
In the main text, we perform renormalization group (RG) transformation by coarse-grain $b=2$ sites (see in particular around \cref{eq:RG-Gamma,eq:RG-h}).
Here, we show that whether or not the RG flow is chaotic does not depend on the value of $b\in \mathbb{Z}_{\geq 2}$.

\subsection{Mathematical definition of Lyapunov exponents}
The defining character of chaos is sensitivity to the initial choice of parameters~\cite{strogatz2024nonlinear}.
Such sensitivity is quantified by a Lyapunov exponent.
Consider a map $x_{n+1} = f(x_{n})$.
For two initial conditions $x_{0}$ and $y_{0} \coloneqq x_{0} + \delta_{0}$, if $\delta_{n} \coloneqq x_{n} - y_{n}$ satisfies $|\delta_{n}|\sim |\delta_{0}| \mathrm{e}^{n \lambda}$,
$\lambda$ is called the Lyapunov exponent of $f$, which is defined as~\cite{strogatz2024nonlinear}
\begin{equation}\label{eq:lambda-def}
  \lambda \coloneqq \lim_{n\to \infty} \frac{1}{n}\sum_{i=0}^{n-1} \log |f'(x_{i})|.
\end{equation}
If the map $f(x)$ is chaotic, then $\lambda>0$.
Below, we show that the RG equation under consideration has a positive Lyapunov exponent when $h>h_{\mathrm{c}}(\Gamma)$.

\subsection{Calculation of the Lyapunov exponent for a generic block-spin transformation}
For a generic $b$, the RG equation under consideration is given by  $K^b[\bm{g}']\propto K[\bm{g}]$ [\cref{eq:K-RG} in the main text] with
\begin{equation}\label{aeq:Kg-def}
  K[\bm{g}=(\Gamma, h)] \coloneqq \sqrt{U}M_{m=+1}\sqrt{U} = \mathrm{e}^{\mathrm{i}h/2 \sigma_{z}} \frac{\exp\left( \Gamma \sigma_{x} \right) }{\sqrt{2 \cosh  2\Gamma}}\mathrm{e}^{\mathrm{i}h/2 \sigma_{z}},
\end{equation}
where $\Gamma >0$ represents the measurement strength and $h \in [0,\pi/2]$.
In this section, we show that this RG equation is chaotic (i.e. it has a positive Lyapunov exponent) for $|h|>h_{\mathrm{c}}(\Gamma)$, irrespective of the value of $b$.
To this end, we note that $K$ has the form
\begin{equation}\label{eq:logK}
  K = \sqrt{\frac{1}{2 \cosh  2\Gamma}}\exp \left[ \mathrm{i}\theta R \right],
\end{equation}
where 
\begin{equation}\label{aeq:R}
  R \coloneqq \frac{-\mathrm{i}}{\sqrt{|\alpha|}}\sigma_{x} + \sqrt{1+\frac{1}{|\alpha|}}\sigma_{z}
\end{equation}
and
\begin{equation}\label{aeq:theta-def}
  \theta \coloneqq -\mathrm{i}\log \left[ \cosh \Gamma \left(\cos h + \mathrm{i}\sqrt{\sin^2 h - \tanh^2 \Gamma}\right) \right],
\end{equation}
which is real in the PT-broken phase.
Here,
\begin{equation}\label{aeq:alpha}
  \alpha\coloneqq 1- \coth^2 \Gamma \sin^2 h
\end{equation}
is an RG invariant.
Now, from \cref{eq:logK}, the RG equation can be recast as
\begin{equation}\label{eq:RG-theta-b}
  \theta' = b \theta \mod 2\pi.
\end{equation}
By defining $x\coloneqq \sin^2 \theta$, this is equivalent to
\begin{equation}\label{eq:f-r}
  x' = f(x), \quad\text{where}\quad f(x) \coloneqq \sin^2 (b \arcsin \sqrt{x}).
\end{equation}
Note that we obtain the logistic map [\cref{eq:logistic} in the main text] if we substitute $b=2$ in \cref{eq:f-r}.

We repeatedly apply this transformation.
Let us consider the sequence $x_{0}, x_{1}, x_{2},\cdots,$ where
\begin{equation}
  x_{0}\eqqcolon \sin^2 \theta,\quad 
  x_{l} \coloneqq f^{l}(x_{0}) = \sin^2 (b^{l}\theta).
\end{equation}
Here, we make the following assumption for $\theta$:
\begin{equation}\label{eq:assumption}
  \sin (2 b^{n}\theta) \neq 0, \quad \forall n\in \mathbb{Z}_{\geq 0}.
\end{equation}
Since $\theta$ is determined by the system parameters $\Gamma$ and $h$ as in \cref{aeq:theta-def}, this condition is generally satisfied.
Then, noting that
\begin{equation}
  f'(x) = b \frac{\sin (2b \arcsin \sqrt{x})}{2\sqrt{x(1-x)}},
  \quad\text{hence}\quad
  f'(x_{l}) = b \frac{\sin (2b^{l+1}\theta)}{\sin (2b^{l}\theta)},
\end{equation}
we have
\begin{align}
  \lambda 
  &= \lim_{n\to \infty} \frac{1}{n}\sum_{l=0}^{n-1}\left( \log  b + \log \left| \frac{\sin (2 b^{l+1}\theta)}{\sin (2b^{l}\theta)} \right|  \right)  \\
  &= \log b + \lim_{n\to \infty} \frac{1}{n}\log  \left| \frac{\sin (2b^{n}\theta)}{\sin (2b^{n-1}\theta)}\cdot \frac{\sin (2b^{n-1}\theta)}{\sin (2b^{n-2}\theta)} \cdots \frac{\sin (2b\theta)}{\sin (2\theta)} \right|  \\
  &= \log b + \lim_{n\to \infty}\frac{1}{n}\log \left| \frac{\sin (2b^{n}\theta)}{\sin (2\theta)} \right|.
  \label{eq:lambda-temp}
\end{align}
The assumption $x_{0} = \sin^2 \theta \neq 0,1$ indicates that $|\sin (2\theta)| \neq 0$.
Together with \cref{eq:assumption}, we can ignore the second term in Eq.~\eqref{eq:lambda-temp} and finally obtain
\begin{equation}
    \lambda = \log  b > 0.  
\end{equation}
It follows that, as long as $x_{0}\in (0, 1)$ (i.e., in PT-broken phase), a generic block-spin transformation with $b$ sites is chaotic.

\section{Details on the chaotic and non-chaotic phases of the RG flow}
In the main text, we consider the RG transformation of a single-qubit nonunitary dynamics, and argue that the resulting flow in \cref{eq:logistic} has a chaotic transition at $h=h_{\mathrm{c}}(\Gamma)$.
Here, we give a detailed account of the chaotic phase.

Recall that the defining features of chaos are given by the following three conditions~\cite{strogatz2024nonlinear}:
\renewcommand{\theenumi}{\Roman{enumi}}
\begin{enumerate} 
  \item The transformation rule is \textit{deterministic}, i.e., the irregular behavior arises from the nonlinearity rather than random inputs.
    \label{item:deterministic}
  \item The dynamics has an exponential sensitivity to a change in initial parameters, i.e., it has a positive Lyapunov exponent $\lambda \gneq 0$. \label{item:sensitivity}
  \item Most trajectories do not converge to fixed points or (quasi-)periodic orbits as the transformation is repeatedly applied.
    \label{item:nonperiodic}
\end{enumerate}
As long as we consider a postselected quantum trajectory, the RG transformations [\cref{eq:K-RG,eq:Phi-RG} in the main text] are deterministic, so that the condition~\ref{item:deterministic} is satisfied.
In the rest of this section, we therefore clarify the physical origins of the remaining conditions~\ref{item:sensitivity} and~\ref{item:nonperiodic}.

\subsubsection{Condition~\ref{item:sensitivity}: Sensitivity to microscopic system parameters}
The physical origin of the sensitivity in the PT-broken phase can be attributed to persistent oscillations caused by the unitary evolution,
which dominates the effects of quantum measurements.
More precisely, in the PT-broken phase, the state does not converge to a steady state in the course of dynamics, and keeps oscillating.
Consequently, a slight variation in the system parameters, which affects the oscillation periods, is amplified over time and leads to an $O(1)$ difference  in the state at late times.
This difference signals the sensitivity to microscopic details.

To quantify such a sensitivity in the PT-broken phase, let us consider two systems prepared in the same initial state $\ket{\psi_{0}}$, and let them evolve separately with Kraus operators with slightly different sets of system parameters $(\Gamma_{1}, h_{1})$ and $(\Gamma_{2},h_{2})$, both of which belong to the same phase (PT-broken or unbroken).
We then consider how the distance between the two states changes in time.

As a distance measure, we consider the trace distance.
Because the states remain pure in the dynamics, it reads
\begin{equation}\label{eq:trace-distance}
  D(\ket{\psi_{1}(N)}, \ket{\psi_{2}(N)}) = \sqrt{1-\left| \braket{\psi_{1}(N)|\psi_{2}(N)} \right|^2},
\end{equation}
where the two states $\ket{\psi_{i}(N)}$ ($i=1, 2$) at time $N$ are given by
\begin{equation}\label{eq:psi12-def}
  \ket{\psi_{i}(N)} \coloneqq \frac{K^{N}(\Gamma_{i}, h_{i})\ket{\psi_{0}}}{\|K^{N}(\Gamma_{i}, h_{i})\ket{\psi_{0}}\|}, \quad i = 1,2.
\end{equation}
Note that $D(\ket{\psi_{1}}, \ket{\psi_{2}}) = 0$ ($D(\ket{\psi_{1}}, \ket{\psi_{2}}) = 1$) if $\ket{\psi_{1}}=\ket{\psi_{2}}$ ($\ket{\psi_{1}} \perp \ket{\psi_{2}}$).

\begin{figure}[tpb]
  \centering
  \includegraphics[width=0.8\textwidth]{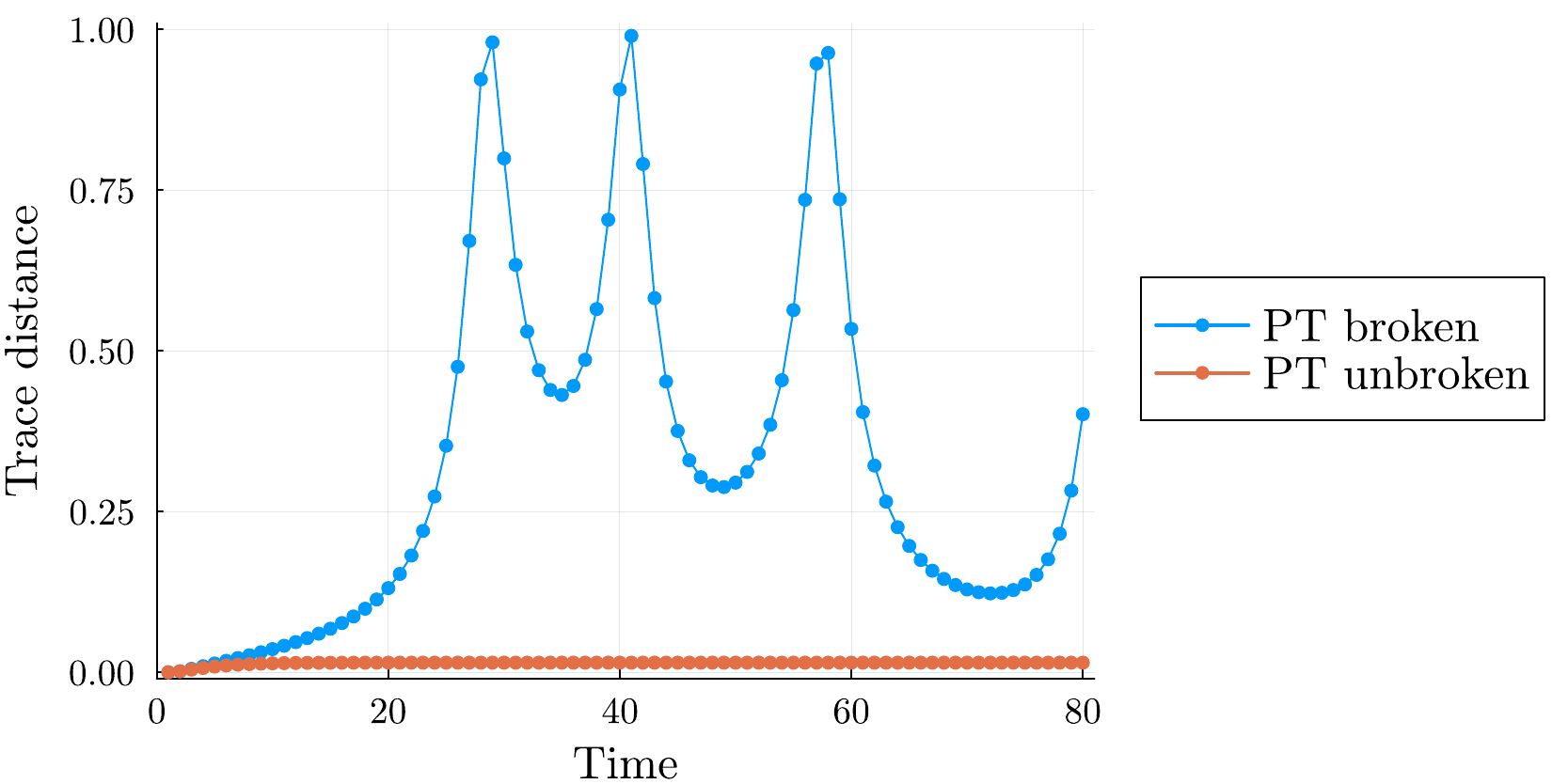}
  \caption{
    Discrete-time evolution of the trance distance~\eqref{eq:trace-distance} between two states with different sets of parameters. 
    In the PT-broken (chaotic) phase of $K$, the trace distance between the two states $\ket{\psi_{1}}$ and $\ket{\psi_{2}}$ in \cref{eq:psi12-def} become $O(1)$ in finite time,
    while in the PT-unbroken (non-chaotic) phase it remains close to $0$ at all time.
    Here we have chosen $(\Gamma_{1}, h_{1}) = (0.3, 0.1\pi)$, $(\Gamma_{2}, h_{2})=(0.31, 0.1\pi)$ in the PT-broken phase, while $(\Gamma_{1}, h_{1}) = (0.4, 0.1\pi)$, $(\Gamma_{2}, h_{2})=(0.41, 0.1\pi)$ in the PT-unbroken phase.
    The initial state is set to be $\ket{\psi_{0}}= (1/\sqrt{2},1/\sqrt{2})^{\mathsf{T}}$.
  }
  \label{fig:trace-distance}
\end{figure}

\Cref{fig:trace-distance} shows typical behavior of the trace distance.
When both $(\Gamma_{1}, h_{1})$ and $(\Gamma_{2}, h_{2})$ belong to the PT-broken phase (i.e., accompanying chaotic RG flow), the trace distance is $O(1)$ in finite time.
By contrast, in the PT-unbroken (non-chaotic) phase, it stays close to $0$ at all times.
These observations indicate that the long-time behavior is sensitive to a small change in system parameters in the PT-broken phase, while
it is insensitive in the PT-unbroken phase.

As another measure of the sensitivity, let us consider the expectation value of an observable to be measured.
To this end, let
\begin{equation}\label{eq:sigmax-N}
  \braket{\sigma_{x}}(N; \Gamma, h) \coloneqq \frac{\braket{\psi_{0}|K^{\dagger N}(\Gamma, h)\sigma_{x}K^{N}(\Gamma,h)|\psi_{0}}}{\braket{\psi_{0}|K^{\dagger N}(\Gamma, h)K^{N}(\Gamma,h)|\psi_{0}}}
\end{equation}
be the expectation value of $\sigma_{x}$ at time $N$ with respect to the state evolved with parameters $(\Gamma, h)$.

\begin{figure}[tpb]
  \centering
  \includegraphics[width=.8\textwidth]{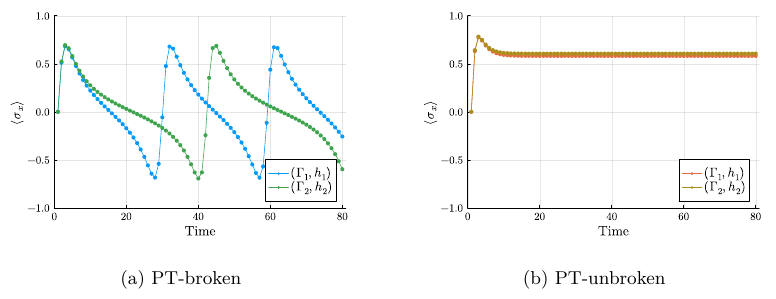}
  \caption{
    Time evolution of the expectation value of $\sigma_{x}$ in each phase.
    Here, $(\Gamma_{1,2},h_{1,2})$ are chosen as the same values in \cref{fig:trace-distance}.
    (a) In the PT-broken phase, tiny differences between
    $(\Gamma_{1},h_{1})$ and $(\Gamma_{2},h_{2})$ are amplified by unitary oscillation,
    resulting in an $O(1)$ difference in $\braket{\sigma_{x}}$ at late times.
    (b) In the PT-unbroken phase, states converge to steady
    states determined by each parameter set, and consequently there is no $O(1)$ difference in $\braket{\sigma_{x}}$.
  }
  \label{fig:sigmax}
\end{figure}

\Cref{eq:logK} shows that $\braket{\sigma_{x}}(N; \Gamma, h)$ depends on $N$ only through a variable $\mathrm{e}^{\mathrm{i}N\theta}$.
Indeed, since $R^2=\mathbf{1}$, one has
\begin{align}
  K^{N}
  &=\left(\frac{2}{\sinh 2\Gamma}\right)^{N/2}\left[ \cos (N\theta) + \mathrm{i}R \sin (N\theta) \right] \nonumber \\
  &\propto \frac{\mathbf{1}+R}{2}\mathrm{e}^{2\mathrm{i}N\theta} + \frac{\mathbf{1}-R}{2},
\end{align}
where we neglect the proportionality factor since it is canceled in forming the ratio as in \cref{eq:sigmax-N}.
Let us regard \cref{eq:sigmax-N} as a function of $\mathrm{e}^{\mathrm{i}N\theta}$ and put $\braket{\sigma_{x}}(N; \Gamma,h) \eqqcolon f(\mathrm{e}^{\mathrm{i}N\theta})$.
Then, if we change $\theta$ as $\theta \to \theta + \delta \theta$, the corresponding change in $\braket{\sigma_{x}}$ reads
\begin{equation}
  \delta \braket{\sigma_{x}} = \delta \theta \cdot f'(\mathrm{e}^{\mathrm{i}N\theta})\cdot \mathrm{i}N\mathrm{e}^{\mathrm{i}N\theta} + O(\delta\theta^2).
\end{equation}
Hence, $\delta\braket{\sigma_{x}}=O(1)$ at $N=O(\delta \theta^{-1})$.
This means that, no matter how small the change $\delta \theta$ in the system parameters is, the difference in $\braket{\sigma_{x}}$ is amplified to $O(1)$ at a sufficiently late time $N$, as long as $f$ is not a constant function.
The time evolution of \cref{eq:sigmax-N} is shown in \cref{fig:sigmax}.

\subsubsection{Condition~\ref{item:nonperiodic}: Absence of (quasi-)periodic orbits}
Physically, the aperiodicity in the RG transformation arises from the incommensurability between the discrete-time unit (which we take as unity) and the oscillation period.
Recall that the RG transformation is equivalent to \cref{eq:RG-theta-b}.
Importantly, $b$ is an integer since we consider the stroboscopic dynamics;
the coarse-graining scale is automatically set by the discrete-time unit.
On the other hand, the variable $\theta$, as determined by the system parameters in \cref{aeq:theta-def}, is in general incommensurate with integer multiples of $\pi$.
This fact, $\theta \notin \pi \mathbb{Q}$, forbids the RG transformation~\eqref{eq:RG-theta-b} to have any (quasi-)periodic orbits.
Thus, discreteness in the dynamics is essential for the emergence of chaotic RG.

\section{Duality to the classical non-Hermitian Ising model}
\subsection{Mapping to a classical non-Hermitian Ising chain}
In this section, we show that the measurement model of a single qubit in \cref{aeq:Kg-def} can be mapped to the transfer-matrix evolution of the classical non-Hermitian Ising model in one dimension. 
Let us define
\begin{equation}\label{aeq:J}
  J \coloneqq -\frac{1}{2}\log \tanh \Gamma \geq 0,
\end{equation}
so that
\begin{equation}\label{aeq:KT}
  K = c \mathrm{e}^{\mathrm{i}h/2 \sigma_{z}}  \begin{pmatrix} \mathrm{e}^{J} & \mathrm{e}^{-J} \\ \mathrm{e}^{-J} & \mathrm{e}^{J} \end{pmatrix} \mathrm{e}^{\mathrm{i}h/2 \sigma_{z}} = c\begin{pmatrix} \mathrm{e}^{J+\mathrm{i}h} & \mathrm{e}^{-J} \\ \mathrm{e}^{-J} & \mathrm{e}^{J + \mathrm{i}h} \end{pmatrix} \eqqcolon cT,
\end{equation}
where
\begin{equation}\label{aeq:c-def}
  c \coloneqq \frac{1}{\sqrt{2(\mathrm{e}^{2J} + \mathrm{e}^{-2J})}}
\end{equation}
is a proportionality factor.
The matrix $T$ is nothing but the transfer matrix of the one-dimensional classical non-Hermitian Ising model with Hamiltonian
\begin{equation}\label{aeq:HIsing}
  -\beta H_{\text{Ising}} = J \sum_{i}S_{i}S_{i+1} + \mathrm{i}h \sum_{i}S_{i}, \quad S_{i}\in \left\{ \pm 1 \right\},
\end{equation}
where $\beta$ is the inverse temperature.

The essence of the mapping lies in identifying the Kraus operator $K$ with the transfer matrix $T$ as in \cref{aeq:KT}.
While $K$ is a mapping in the real-time direction of the quantum system, $T$ is a mapping in the real-space direction (i.e., to a neighboring site) of the corresponding classical non-Hermitian Ising model.
The two systems become equivalent by exchanging (real-)time with space.

Notably, the magnetic field $\mathrm{i}h$ associated to the classical model is pure imaginary.
In this case, there exists a finite-temperature phase transition even in one dimension as discovered by Lee and Yang~\cite{lee1952statistical};
this is because the partition function $Z_{\text{Ising}}\coloneqq \sum_{\left\{ S_{i} \right\}}\mathrm{e}^{-\beta H_{\text{Ising}}}$, which takes \textit{complex} values due to non-Hermiticity, vanishes at the transition point, thereby giving rise to a divergence in the free-energy density $f_{\text{Ising}} = -\frac{1}{\beta L}\log Z_{\text{Ising}}$.

The Yang-Lee transition can be interpreted as a PT symmetry breaking transition of the transfer matrix $T$~\cite{matsumoto2022embedding}.
On the other hand, the chaotic transition of the RG flow considered in the main text is also due to the PT-transition of $K$.
Thus, from \cref{aeq:KT}, the two transitions are equivalent:
they both take place at $h = h_{\mathrm{c}}(\Gamma) \coloneqq \arcsin \tanh\Gamma = \arcsin \mathrm{e}^{-2J}$~\cite{fisher1980yang}.

A few remarks are in order.
First, the correspondence presented in this section differs from the conventional quantum-classical correspondence~\cite{suzuki1971relationship, suzuki1976relationship,  fradkin1978order, Sachdev2011quantum, kogut1979introduction}, in which quantum \textit{imaginary} time, rather than real time, is related to classical real space. See Refs.~\cite{matsumoto2022embedding, gao2024experimental, xu2025characterizing} for application of the quantum-classical correspondence to the Yang-Lee problem.
Second, although the classical non-Hermitian Ising models have attracted growing interest due to their nonunitary critical phenomena~\cite{fisher1978yang, Ashida_2020, cardy2024yang}, their experimental realization is generally not straightforward~\cite{wei2012lee, peng2015experimental, gao2024experimental, shen2023proposal, lu2025dynamical}.
The method presented here enables one to map such non-Hermitian classical spin chains onto repeated-measurement dynamics of a finite-level (i.e., a zero-dimensional) system, which can be experimentally probed as seen in \cref{asec:procedure}.
Compared with existing approaches~\cite{wei2012lee, matsumoto2022embedding}, our method, based on discrete-time evolution, is capable of faithfully implementing the discrete lattice structure of the classical model. 
Moreover, as shown in \cref{asec:correlation}, it allows for the evaluation of multi-point correlation functions by utilizing distinct quantum trajectories.

\subsection{Procedure to obtain classical non-Hermitian partition functions}
\label{asec:procedure}
By using \cref{aeq:KT}, the partition function for the model in \cref{aeq:HIsing} can be experimentally measured using a single-qubit measurement scheme.
To show this, we first note that the partition function $Z_{\text{Ising}}$ can be written in terms of the transfer matrix $T$ as
\begin{equation}\label{aeq:Z-general}
  Z_{\text{Ising}} = \sum_{\left\{ S_{i} \right\}} \mathrm{e}^{-\beta H_{\text{Ising}}} = \Tr (X T^{L}),
\end{equation}
where $L$ is the size of the system and the matrix $X$ encodes the boundary conditions.
For example, $X=\mathbf{1}$ represents the periodic boundary conditions.
To obtain $Z_{\text{Ising}}$ experimentally, special care must be taken with the boundary conditions.
We consider three prototypical cases below.

\subsubsection{Fixed boundary conditions}
For fixed boundary conditions, we take $X = |S_{1})(S_{L}|$ in \cref{aeq:Z-general}, where $|S_{1, L}) \in \left\{ \mid\uparrow), \mid\downarrow) \right\}$ for $\mid\uparrow) \coloneqq (1,0)^{\mathsf{T}}$ and $\mid\downarrow) \coloneqq (0,1)^{\mathsf{T}}$ being classical spin state vectors.
(Hereafter, we will use the notation $|\cdot)$ for classical spin vectors, and $\ket{\cdot}$ for quantum states.)
In this case, by setting $\ket{\psi_{\mathrm{i}}}\propto |S_{1})$ and $\ket{\psi_{\mathrm{f}}}\propto |S_{L})$ in an appropriate basis, we have
\begin{equation}\label{aeq:ZIsing}
  Z_{\text{Ising}} = (S_{L}|T^{L}|S_{1}) = c^{-L}\braket{\psi_{\mathrm{f}}|K^{L}|\psi_{\mathrm{i}}},
\end{equation}
where $c$ is defined in \cref{aeq:c-def}.
An experimental scheme to probe the right-hand side is as follows.
First, a single qubit is prepared in an initial state $\ket{\psi_{\mathrm{i}}}$. The Kraus operator $K$ is then applied $L$ times, yielding the normalized state $K^{L}\ket{\psi_{\mathrm{i}}}/\|K^{L}\ket{\psi_{\mathrm{i}}}\|$. This procedure can be implemented via postselection with success probability $p \coloneqq \|K^{L}\ket{\psi_{\mathrm{i}}}\|^2$.
Finally, the transition probability to a final state $\ket{\psi_{\mathrm{f}}}$ is measured, giving $|\braket{\psi_{\mathrm{f}}|K^{L}\ket{\psi_{\mathrm{i}}}}|^2/p$. From this, $|Z_{\text{Ising}}|$ is obtained up to known coefficients.
As noted around \cref{eq:Mm} in the main text, the Kraus operator $K = K_{m=+1}$ can be implemented via a weak measurement of the spin by using, e.g., ancillary qubits as discussed in \cref{asec:ancilla}.

\subsubsection{Open boundary conditions}
For open boundary conditions, the partition function~\eqref{aeq:Z-general} reads
\begin{equation}
  Z_{\text{Ising}} = \sum_{S, S' \in \left\{ \uparrow, \downarrow \right\}} (S|T^{L}|S').
\end{equation}
Since $\sum_{S\in \left\{ \uparrow, \downarrow \right\}} |S) = (1,1)^{\mathsf{T}}$, one can simply take
\begin{equation}
  \ket{\psi_{\mathrm{i}}} = \ket{\psi_{\mathrm{f}}} = \frac{1}{\sqrt{2}} \left( 1, 1 \right)^{\mathsf{T}}
\end{equation}
and apply a procedure similar to that in the previous section.

\subsubsection{Periodic boundary conditions}
For periodic boundary conditions, the partition function~\eqref{aeq:Z-general} reads
\begin{equation}\label{aeq:ZJ-PBC}
  Z_{\text{Ising}} = \Tr\left( T^{L} \right) = \sum_{S \in \left\{ \uparrow, \downarrow \right\}} (S|T^{L}|S).
\end{equation}
One can use the following two approaches to obtain $Z_{\text{Ising}}$.

(i) The first approach is to measure
\begin{equation}
  Z_{S}\coloneqq (S|T^{L}|S) \propto \braket{S|K^{L}|S}
\end{equation}
for each $S\in \left\{ \uparrow, \downarrow \right\}$ including the complex phases by using quantum state tomography~\cite{d2003quantum}, and calculate \cref{aeq:ZJ-PBC} by $Z_{\text{Ising}} = \sum_{S}Z_{S}$.

(ii) Alternatively, we prepare two maximally entangled qubits in an initial state
\begin{equation}
  \ket{\Omega}\coloneqq \frac{1}{\sqrt{2}}\sum_{S\in \left\{ \uparrow, \downarrow \right\}} \ket{S} \otimes \ket{S},
\end{equation}
where $\ket{\uparrow}=(1,0)^{\mathsf{T}}$ and $\ket{\downarrow}=(0,1)^{\mathsf{T}}$ in a given basis.
Regarding the first qubit as the system, we evolve it using $K$ in \cref{aeq:Kg-def} and project the total state onto $\ket{\Omega}$.
The probability of the whole process is given by
\begin{equation}\label{aeq:p-bell}
  \left| \bra{\Omega} \frac{1}{\sqrt{2}}\sum_{S\in \left\{ \uparrow, \downarrow \right\}} \left(K^{L}\ket{S}\right) \otimes \ket{S} \right|^2  = \frac{1}{4}|\Tr \left( K^{L} \right) |^2.
\end{equation}
Because of \cref{aeq:KT}, \cref{aeq:p-bell} is proportional to $|Z_{\text{Ising}}|^2$.

\subsection{Correlation functions}
\label{asec:correlation}
Above, we postselect a quantum trajectory in which all the measurement outcomes are $+1$ and discard all other trajectories (see also \cref{eq:Mm} in the main text).
However, we can use some of these discarded trajectories to compute the correlation functions of the classical Ising model.

Let us note that
\begin{equation}\label{aeq:K-pm}
  K = K_{+} = \sigma_{z}K_{-}\sigma_{z},
\end{equation}
where $K_{+} = K_{m=+1}$ and $K_{-} = K_{m=-1}$ are defined below \cref{eq:Mm} in the main text.
Thus, for example, a quantum trajectory of the form
\begin{equation}
  \bm{m}_{1} = (\underbrace{1,\cdots,1}_{L-l-n}, \underbrace{-1,-1,\cdots,-1}_{n}, \underbrace{1,\cdots, 1}_{l})
\end{equation}
satisfies
\begin{equation}
  K_{\bm{m}_{1}} = K_{+}^{L-l-n}\sigma_{z}K_{+}^{n}\sigma_{z}K_{+}^{l},
\end{equation}
where $K_{\bm{m}_{1}}\coloneqq K_{(m_{1})_{L}}K_{(m_{1})_{L-1}}\cdots K_{(m_{1})_{1}}$.
If we define  $\bm{m}_{0}\coloneqq (+1, \cdots, +1)$, the ratio
\begin{equation}
  \frac{|\braket{\psi_{\mathrm{f}}|K_{\bm{m}_{1}}|\psi_{\mathrm{i}}}|}{|\braket{\psi_{\mathrm{f}}|K_{\bm{m}_{0}}|\psi_{\mathrm{i}}}|}
  = \frac{\left| \braket{\psi_{\mathrm{f}} | K_{+}^{L-l-n}\sigma_{z}K_{+}^{n}\sigma_{z}K_{+}^{l} | \psi_{\mathrm{i}}} \right| }{\left| \braket{\psi_{\mathrm{f}}|K_{+}^{L} | \psi_{\mathrm{i}}}  \right| }
\end{equation}
coincides with the absolute value of the two-point correlation function for the classical Ising chain,
\begin{equation}
  \left| \braket{S_{l+n}S_{l}} \right|  \coloneqq \frac{\left| \left(S_{L} | T^{L-l-n}\sigma_{z}T^{n}\sigma_{z}T^{l} | S_{1}\right)  \right| }{\left| \left(S_{L} | T^{L} | S_{1}\right)  \right| }.
\end{equation}

Likewise, if one wishes to measure the magnetization of the classical model, one can choose a quantum trajectory
\begin{equation}
  \bm{m}_{2} \coloneqq (\underbrace{-1,\cdots,-1}_{L-n},\underbrace{1,\cdots,1}_{n})
\end{equation}
as well as the modified boundary condition $\bra{\psi_{\mathrm{f}}}\sigma_{z}$, so that
\begin{equation}
  \frac{\tilde{Z}[\bm{m}_{2}]}{\tilde{Z}[\bm{m}_{0}]} = \frac{\left| \braket{\psi_{\mathrm{f}} | K_{+}^{L-n}\sigma_{z}K_{+}^{n} | \psi_{\mathrm{i}}} \right| }{\left| \braket{\psi_{\mathrm{f}}|K_{+}^{L} | \psi_{\mathrm{i}}}  \right| }
   = |\braket{S_{n}}|.
\end{equation}

\subsection{Emergence of the Yang-Lee universality class in measurement dynamics}
\subsubsection{Magnetization}
Here, we demonstrate the emergence of the one-dimensional Yang-Lee universality class in the measurement process.
Let us first recall the expression for the local magnetization of the classical Ising model~\eqref{aeq:HIsing},
\begin{equation}\label{aeq:m}
  m = \frac{\Tr (T^{L}\sigma_{z})}{\Tr T^{L}},
\end{equation}
which diverges at the Yang-Lee edge singularity $h = h_{\mathrm{c}} = \arcsin \mathrm{e}^{-2J}$ as
\begin{equation}\label{aeq:YL-universality}
  \lim_{h\nearrow h_{\mathrm{c}}}\lim_{L \to \infty}m \propto (h_{\mathrm{c}} - h)^{\sigma}, \quad \sigma = -\frac{1}{2},
\end{equation}
where $h_{\mathrm{c}} = \arcsin \mathrm{e}^{-2J}$. 
Here, $h \nearrow h_{\mathrm{c}}$ denotes the limit in which $h$ approaches $h_{\mathrm{c}}$ from below, i.e., from the PT-unbroken phase.
The critical exponent $\sigma$ characterizes the Yang-Lee universality class in one dimension~\cite{fisher1980yang}.

We show, by explicit calculation, that the following quantity in the measurement dynamics exhibits the same critical behavior:
\begin{equation}\label{aeq:sz}
  \tilde{m} := \lim_{h \nearrow h_{\mathrm{c}}}\lim_{L_{1,2} \to \infty} \frac{\Tr\left( K^{L_{1}}\sigma_{z}K^{L_{2}}\rho K^{\dagger}{}^{L_{1}+L_{2}} \right) }{\Tr \left( K^{L_{1}+L_{2}} \rho K^{\dagger}{}^{L_{1} + L_{2}} \right) },
\end{equation}
where $\rho$ is some initial state and $K$ is defined in \cref{aeq:KT}.
For a positive integer $N$, one can show by mathematical induction that
\begin{equation}\label{aeq:RN-1}
  K^{N}\coloneqq 
  \left[ \frac{1}{\sqrt{4 \cosh  2J}} \begin{pmatrix} \mathrm{e}^{J+\mathrm{i}h} & \mathrm{e}^{-J} \\ \mathrm{e}^{-J} & \mathrm{e}^{J-\mathrm{i}h} \end{pmatrix}\right]^{N}
  = \left( \frac{\tanh 2J}{2} \right)^{N/2} \begin{pmatrix}
    A_{N} & B_{N} \\ B_{N} & A_{N}^{*}
  \end{pmatrix},
\end{equation}
where
\begin{align}
  A_{N} &\coloneqq   \cosh N\varphi + \mathrm{i}\frac{\sinh N\varphi}{\sinh \varphi} \sqrt{\frac{\mathrm{e}^{-2J}}{\mathrm{e}^{2J}-\mathrm{e}^{-2J}} - \sinh^2 \varphi} = \cosh N\varphi + \mathrm{i}\sinh N\varphi \sqrt{-\left( 1-\alpha^{-1} \right) },
  \\
  B_{N} &\coloneqq \frac{\sinh  N\varphi}{\sinh \varphi}\sqrt{\frac{\mathrm{e}^{-2J}}{\mathrm{e}^{2J}-\mathrm{e}^{-2J}}}
  = \sinh N\varphi \cdot \sqrt{\alpha^{-1}}.
  \label{aeq:B-def}
\end{align}
Here, $\alpha = 1-\mathrm{e}^{4J}\sin^2 h$ is the same as the one in \cref{aeq:alpha} under \cref{aeq:J}, and
\begin{equation}
  \varphi \coloneqq \log  \frac{\mathrm{e}^{J}\cos h + \sqrt{-\mathrm{e}^{2J}\sin^2 h + \mathrm{e}^{-2J}}}{\sqrt{\mathrm{e}^{2J}-\mathrm{e}^{-2J}}}
\end{equation}
is a real parameter when $h<h_{\mathrm{c}}$.
For $N\gg 1$, we can approximate $\cosh N\varphi$ and $\sinh N\varphi$ as $\mathrm{e}^{N\varphi}/2$.
Hence, \cref{aeq:RN-1} simplifies as
\begin{equation}
  K^{N} \sim 
  \left( \frac{\tanh 2J}{2} \right)^{N/2} \frac{\mathrm{e}^{N\varphi}}{2\sqrt{\alpha}}\begin{pmatrix}
    \sqrt{\alpha}+\mathrm{i}\sqrt{1-\alpha} & 1 \\ 1 & \sqrt{\alpha}-\mathrm{i}\sqrt{1-\alpha}
  \end{pmatrix}, \quad N \gg 1.
\end{equation}
Using this formula, the leading contribution to the denominator of \cref{aeq:sz} reads
\begin{equation}\label{aeq:denom}
  \Tr \left( K^{L_{1}+L_{2}} \rho K^{\dagger}{}^{L_{1} + L_{2}} \right) 
  \sim \left( \frac{\tanh 2J}{2} \right)^{L} \frac{\mathrm{e}^{2L\varphi}}{2\alpha}c, \quad L_{1}, L_{2}\gg 1,
\end{equation}
where $L\coloneqq L_{1}+L_{2}$ and
\begin{equation}
  c\coloneqq \Tr \left[ \begin{pmatrix} 1 & \sqrt{\alpha}-\mathrm{i}\sqrt{1-\alpha} \\ \sqrt{\alpha}+\mathrm{i}\sqrt{1-\alpha} & 1 \end{pmatrix} \rho \right] = O(1)
\end{equation}
is a nonuniversal constant.
(Note that $\alpha \searrow 0$ as $h\nearrow h_{\mathrm{c}}$.)
Likewise, the leading contribution to the numerator of \cref{aeq:sz} reads
\begin{equation}\label{aeq:num}
  \Tr\left( K^{L_{1}}\sigma_{z}K^{L_{2}}\rho K^{\dagger}{}^{L_{1}+L_{2}} \right) 
  \sim \left( \frac{\tanh 2J}{2} \right)^{L} \frac{\mathrm{e}^{2L\varphi}}{(2\sqrt{\alpha})^3} c',
  \quad L_{1}, L_{2}\gg 1,
\end{equation}
where
\begin{equation}
  c'\coloneqq   \Tr\left[ 4 \begin{pmatrix} \mathrm{i}\sqrt{1-\alpha} & 1-\alpha+\mathrm{i}\sqrt{\alpha(1-\alpha)} \\ -1+\alpha+\mathrm{i}\sqrt{\alpha(1-\alpha)} & \mathrm{i}\sqrt{1-\alpha} \end{pmatrix} \rho \right] = O(1)
\end{equation}
is again a constant.
Substituting \cref{aeq:denom,aeq:num} into \cref{aeq:sz}, we obtain the asymptotic behavior
\begin{equation}\label{aeq:mtilde-criticality}
  \tilde{m} \propto \lim_{h \nearrow h_{\mathrm{c}}} \alpha^{-1/2} \propto \lim_{h \nearrow h_{\mathrm{c}}} (h_{\mathrm{c}}-h)^{\sigma},
\end{equation}
where the critical exponent $\sigma = -1/2$ is the same as that in \cref{aeq:YL-universality}.
Thus, we confirm that the critical behavior of $\tilde{m}$ belongs to the Yang-Lee universality class.

\subsubsection{Correlation function}
One can consider other critical exponents in a similar manner.
For example, the two-point correlation function in the classical model behaves as~\cite{fisher1980yang}
\begin{equation}\label{aeq:G-critical}
  G(x) \coloneqq \lim_{L \to \infty}\frac{\Tr(T^{L}\sigma_{z}T^{x}\sigma_{z})}{\Tr T^{L+x}} - m^2 \propto \frac{\mathrm{e}^{-x/\xi}}{(x/\xi)^2} x^{-(d-2+\eta)},
\end{equation}
where $m$ is defined in \cref{aeq:m}, $d=1$ is the spatial dimension, $\eta=-1$ is the anomalous dimension, and $\xi$ is the correlation length which diverges as $\lim_{h \nearrow h_{\mathrm{c}}}\xi \propto (h_{\mathrm{c}}-h)^{-\nu}$ with $\nu = 1/2$.
The corresponding quantity in the measurement dynamics is the two-time correlation function
\begin{equation}\label{aeq:Gtilde}
  \tilde{G}(x) \coloneqq \lim_{L_{1,2} \to \infty} \frac{\Tr (K^{L_{1}}\sigma_{z}K^{x}\sigma_{z}K^{L_{2}}\rho K^{\dagger}{}^{L_{1}+L_{2}+x})}{\Tr \left( K^{L_{1}+L_{2}+x} \rho K^{\dagger}{}^{L_{1}+L_{2}+x} \right) } - \tilde{m}^2.
\end{equation}
Following a similar calculation that led to \cref{aeq:mtilde-criticality}, one can confirm that $\tilde{G}(x)$ also exhibits the same critical behavior in \cref{aeq:G-critical}.

\subsubsection{Other physical quantities}
Although the quantities in \cref{aeq:sz,aeq:Gtilde} exhibit the Yang-Lee critical exponents, it is difficult to probe them experimentally since they do not appear as a consequence of physical time evolutions.
Alternatively, let us consider the following quantity:
\begin{equation}\label{aeq:mp-tilde}
  \tilde{m}' 
  \coloneqq 
  \lim_{h \nearrow h_{\mathrm{c}}} \lim_{L_{1,2}\to \infty} \mu(h,L_{1},L_{2})
  = \lim_{h \nearrow h_{\mathrm{c}}}\lim_{L_{1,2} \to \infty} \frac{\Tr\left( K^{L_{1}}\sigma_{z}K^{L_{2}}\rho K^{\dagger}{}^{L_{2}}\sigma_{z}K^{\dagger}{}^{L_{1}}\right) }{\Tr \left( K^{L_{1}+L_{2}} \rho K^{\dagger}{}^{L_{1} + L_{2}} \right) },
\end{equation}
where $\mu(h,L_{1},L_{2})$ is defined in \cref{eq:mu-def} in the main text.
Due to \cref{aeq:K-pm}, it can also be written as
\begin{equation}
  \tilde{m}' \coloneqq \lim_{h \nearrow h_{\mathrm{c}}}\lim_{L_{1,2} \to \infty} \frac{\Tr\left( K_{-}^{L_{1}}K_{+}^{L_{2}}\rho K_{-}^{\dagger}{}^{L_{2}}K_{+}^{\dagger}{}^{L_{1}}\right) }{\Tr \left( K_{+}^{L_{1}+L_{2}} \rho K_{+}^{\dagger}{}^{L_{1} + L_{2}} \right) }.
\end{equation}
Thus, 
we see that the numerator of this expression corresponds to the probability of a quantum trajectory with outcomes
\begin{equation}
  (\underbrace{-1,\cdots,-1}_{L_{1}},\underbrace{1,\cdots,1}_{L_{2}}).
\end{equation}
On the other hand, calculations analogous to \cref{aeq:denom,aeq:num} yield
\begin{equation}
  \tilde{m}' \propto \lim_{h\nearrow h_{\mathrm{c}}} \alpha^{-1} \propto \lim_{h\nearrow h_{\mathrm{c}}}(h_{\mathrm{c}}-h)^{2\sigma},
  \quad \sigma = -\frac{1}{2}.
\end{equation}
The factor of two in front of $\sigma$ in the exponent originates from the fact that $\sigma_{z}$ appears twice in the definition in \cref{aeq:mp-tilde}.

Likewise, we can consider the following quantity which is a counterpart to $\tilde{G}(x)$ in \cref{aeq:Gtilde}:
\begin{equation}
  \tilde{G}'(x) \coloneqq \lim_{L_{1,2} \to \infty} g(x,L_{1},L_{2}) - \lim_{x \to \infty}\lim_{L_{1,2} \to \infty} g(x,L_{1},L_{2}),
\end{equation}
where
\begin{equation}
  g(x,L_{1},L_{2}) \coloneqq \frac{\Tr (K^{L_{1}}\sigma_{z}K^{x}\sigma_{z}K^{L_{2}}\rho K^{\dagger}{}^{L_{2}}\sigma_{z} K^{\dagger}{}^{x}\sigma_{z}K^{\dagger}{}^{L_{1}})}{\Tr \left( K^{L_{1}+L_{2}+x} \rho K^{\dagger}{}^{L_{1}+L_{2}+x} \right) }.
\end{equation}
The numerator of $g(x,L_{1},L_{2})$ can be obtained by choosing a quantum trajectory with outcomes
\begin{equation}
  (\underbrace{1,\cdots,1}_{L_{1}}, \underbrace{-1,\cdots,-1}_{x}, \underbrace{1,\cdots,1}_{L_{2}}).
\end{equation}
Following a similar calculation, one can show that $\tilde{G}'(x)$ exhibits the following behavior:
\begin{equation}
  \tilde{G}'(x) \propto \left[ \frac{\mathrm{e}^{-x/\xi}}{(x/\xi)^2} \right]^2 x^{-2(d-2+\eta)},
\end{equation}
where $d=1$ and $\eta=-1$.

\section{Implementation of the quantum measurement by ancilla}
\label{asec:ancilla}
In this section, we show that
\begin{equation}
  M_{\pm} 
  = \frac{1}{\sqrt{2 \cosh 2\Gamma}}\mathrm{e}^{\pm \Gamma X}
  = \frac{1}{\sqrt{4 \cosh 2J}}\begin{pmatrix} \mathrm{e}^{J} & \pm \mathrm{e}^{-J} \\ \pm \mathrm{e}^{-J} & \mathrm{e}^{J} \end{pmatrix}
  = \frac{1}{\sqrt{1+(1-\eta)^2}}\begin{pmatrix} 1-\eta/2 & \pm \eta/2 \\ \pm \eta/2 & 1-\eta/2 \end{pmatrix}
\end{equation}
can be implemented experimentally by using an ancilla qubit, where $\sinh 2\Gamma = (\sinh 2J)^{-1}$ and $\eta\coloneqq 2\tanh\Gamma / (1+\tanh\Gamma) = 2/(1+\mathrm{e}^{2J})\in [0,1]$ parametrizes the strength of the measurement.

Using an indirect measurement model, one can show that
\begin{equation}
  M_{m}\rho M_{m}^{\dagger} = \Tr_{\mathrm{A}} \left[ (1\otimes P_{m}) R (\rho\otimes \rho_{\mathrm{a}}) R^{\dagger} (1\otimes P_{m})\right],
\end{equation}
where $\Tr_{\mathrm{A}}$ denotes the partial trace over the ancilla Hilbert space, $P_{\pm}\coloneqq \ket{\pm}\bra{\pm}$ are the projection operators onto the $\pm 1$ eigenspaces of $X$,
\begin{equation}
  R\coloneqq P_{+}\otimes \frac{1}{\sqrt{1+(1-\eta)^2}} \begin{pmatrix} 1-\eta & 1 \\ 1 & -(1-\eta) \end{pmatrix}
  + P_{-}\otimes \frac{1}{\sqrt{1+(1-\eta)^2}}\begin{pmatrix} 1 & 1-\eta \\ 1-\eta & -1 \end{pmatrix}
\end{equation}
is a unitary operator,
and $\rho_{\mathrm{a}}\coloneqq \ket{+}\bra{+}$ is the initial state of the ancilla qubit.
We note that this choice of $P_{m}, R,$ and $\rho_{\mathrm{a}}$ is one example, and that different choices are also possible.
See also Appendix A in Supplementary Material of Ref.~\cite{mochizuki2024measurement}.

\end{document}